\documentclass{article}\usepackage[margin=1in,centering]{geometry}
 \newcommand\ttl {Gacs -- Kucera Theorem} \newcommand\aut {Leonid A.~Levin}
 \usepackage[unicode]{hyperref} \usepackage[utf8]{inputenc}
 \usepackage{microtype}\usepackage{amssymb,mmap}
\begin{document}

\newcommand\edf{{\raisebox{-2pt}{$\,\stackrel{\mbox{\tiny df}}=\,$}}}
\newcommand\W{{\mathbf\Omega}}\newcommand\St{{\mathbf S}}
\newcommand\R{{\mathbf R}}\newcommand\N{{\mathbf N}}\renewcommand\U{{\mathbf U}}
\newcommand\m{{\mathbf m}}\newcommand\M{{\mathbf M}}\newcommand\T{{\mathbf T}}
\newcommand\K{{\mathbf K}}\renewcommand\d{{\mathbf d}}\frenchspacing

\author{\aut\thanks{\url {https://www.cs.bu.edu/fac/Lnd}
 Boston University, CAS, Computer Science department, Boston, MA 02215, USA}}
 \date{}\title{\vspace*{-6pc}\ttl\vspace{-6pt}}\maketitle\vspace{-1pc}

\begin {abstract} Gacs -- Kucera Theorem \cite {g86,ku,L76}, tightened by
Barmpalias, Lewis-Pye \cite{BL}, w.t.t.-reduces each infinite sequence to
a Kolmogorov -- Martin-Lof random one and is broadly used in various Math and
CS areas.\par Its early proofs are somewhat cumbersome, but using some general
concepts yields significant\\ simplification illustrated below.\end{abstract}

\vspace{-6pt}\section {General Terminology}\vspace{-6pt}

 {\bf\em Computably enumerable (c.e.)} functions to $\overline{\R^+}$
 are the supremums of c.e. sets of basic continuous ones.

{\bf\em Dominant} in a convex class $C$ of functions is its c.e. $f\in C$
if all c.e. $g$ in $C$ are $O(f)$.\\ Such is $\sum_ig_i/(i^2{+}i)$ if
$(g_i)$ is a c.e. family of all c.e. functions in weakly compact $C$.

{\bf\em Distributions} on $\St\edf\{0,1\}^*$ are $m\ge0$ with $\sum_xm(x)\le1$.
For $t\in R^+\!$, $\|t\|\edf\lceil\log_2t\rceil{-}1$.

{\bf\em Kolmogorov complexity} $\K(x)$ is $-\|\m(x)\|$,
for a dominant distribution $\m$ on $\St$.

{\bf\em Uniform measure} on infinite sequences $\alpha\in\W\edf\{0,1\}^{\N}$
is $\lambda(x\,\W)\edf2^{-n}$ for $x\in\{0,1\}^n$.

{\bf\em Martin-Lof $\lambda$-test} ${d}(\alpha)$ is $\|\T(\alpha)\|$
for a dominant $\T$ on $\W$ with expectation $\lambda(\T)\le1$.\\
Measure $\tau$ is the $\lambda$-integral of $\T\!$.
(It is $P(\m\otimes\lambda)$ for $P(x,\alpha)\edf x\alpha$.)

\bigskip{\bf\em Partial continuous transforms (PCT)}
 on $\W$ may fail to narrow-down the output\\
to a single sequence, leaving a compact set of eligible results.
 So, their graphs are compact\\ sets $A\subset\W{\times}\W$ with
 $A(\alpha)\edf\{\beta:(\alpha,\beta)\in A\}\ne\emptyset$.
Singleton outputs $\!\{\beta\}$ are interpreted as $\beta\in\W$.

{\bf\em Computable PCT}s have algorithms enumerating
the clopen subsets of $\W^2\setminus A$.

{\bf\em Preimages} $A^{-1}(s)\edf\{\alpha:A(\alpha)\subset s\}$
of all open sets $s\subset\W$ in any PCT $A$ are open.

{\bf\em Closed} $A$ also have closed preimages of all closed $s$.
Such $A$ are {\bf\em $t$-closed} for some~$t$,~i.e.\\ $A^{-1}(s)$ of
clopen $s$ depend only on the first $t(\widehat s)$ bits of $\alpha$.
($\widehat s\subset\St$ is the smallest set with $s=\widehat s\,\W$.)

Below, I assume $t$ computable and prefix-based, i.e.
$t(\widehat s)\ge t(\{x\})$ if $xy\in\widehat s$.

\bigskip{\bf\em Semimeasures} $P(x)\ge P(x0){+}P(x1)\in[0,1]$ are
peculiar probability distributions that PCTs generate from random inputs
as $A(\lambda):x\mapsto\lambda(A^{-1}(x\,\W))$. Any such c.e. $P$ is
generated by a computable $t$-closed PCT\\ if $P(x)$ are binary rationals
of $<t(\{x\})$ bits.\footnote {See \cite{ZL}, proof of Theorem 3.2.}
 {\bf\em Dominant semimeasure $\M$} has values shorter than\\
 $\K(x)$ bits: $\M(x)$ can be so rounded-up\footnote
 {Used (with a $\log\|x\|$ slack) in \cite{L71}, Theorem 13
  (that restates Proposition 3.2 of \cite{ZL}).}
 after adding $\sum_{y\ne\emptyset}\m(xy)$ (to keep $\M(x)\ge\M(x0){+}\M(x1)$).\\
 Thus, $\M$ can be generated from $\lambda$ by a computable
 $t$-closed PCT if $t(\{x\})>\K(x)$. Let $\U(\lambda)=\M$ do that.

\vspace{-6pt}\section {Proof of Gacs -- Kucera Theorem}\vspace{-6pt}

Now, $\exists c\,\U(\tau)<c\,\U(\lambda)$, so the preimages $\U^{-1}(x\,\W)$ of
all $x$ intersect $\rho\edf\{\beta:\T(\beta)\le c\}$. But for closed~$\U$,~the
preimage of $\alpha\,{\in}\,\W$ is the intersection of (non-empty in $\rho$)
closed preimages of its prefixes, so intersects $\rho$, too.

 \end {document}